\documentclass{iopart}

\usepackage{graphicx}

\begin{document}

\title[Electrostatic ion perturbations]{Electrostatic ion perturbations in kinematically
complex shear flows}

\author{Osmanov Z.}

\address{School of Physics, Free University of Tbilisi, 0183, Tbilisi,
Georgia} \ead{z.osmanov@freeuni.edu.ge}

\author{Rogava A.}

\address{Centre for Theoretical Astrophysics, ITP, Ilia State University, 0162-Tbilisi, Georgia}

\author{Poedts S.}

\address{Centre for Plasma Astrophysics, Katholieke Universiteit
Leuven, Celestijnenlaan 200B, bus 2400 B-3001, Belgium}

\begin{abstract}
The scope of the present paper is to determine how ion electrostatic
wave perturbations in plasma flows are influenced by the presence of
a kinematically complex velocity shear. For this purpose we consider
a model based on the following set of physical equations: the
equation of motion, the continuity equation and the Poisson equation
for the electric potential governing the evolution of the system.
After linearizing the equations, we solve them numerically. We find
out that for a variety of specific values of parameters the system
may exhibit quite interesting dynamic behaviour. In particular, we
demonstrate that the system exhibits two different kinds of shear
flow instabilities: (a)~when the wave vectors evolve exponentially,
the ion sound modes become unstable as well; while, (b)~on the other
hand, one can find areas in a parametric space where, when the wave
vectors vary periodically, the physical system is subject to a
strongly pronounced  parametric instability. We also show the
possibility of the generation of beat wave phenomena, characterized
by a noteworthy quasi-periodic temporal behaviour. In the
conclusion, we discuss the possible areas of applications and
further directions of generalization of the presented work.
\end{abstract}

\maketitle

\section{Introduction}

It is well known that plasma flows in different astrophysical,
geophysical and laboratory situations are characterized by spatially
inhomogeneous velocity fields (shear flows) and the presence of this
velocity shear may significantly influence the modes of collective
behaviour fostered in these flows. In most of the cases these flows
are kinematically complex but even in a relatively simple case the
overall dynamics, especially transition of the flow to turbulence,
could be quite problematic \cite{kerswell,schek}. One of the typical
examples of astrophysical flows are helical plasma motions which
occur in extragalactic jets \cite{broder,kharb} and in young stellar
object jets \cite{yso}. It has also been argued that in the solar
atmosphere at least some of macro-spicules are characterized by
`tornado-like' kinematical flow geometry \cite{pm98}.

Obviously, the existence of such shear flows with kinematic
complexity might strongly influence the plasma processes in a number
of realistic astrophysical scenarios. Therefore, it is interesting
to study the behaviour of plasmas being influenced by such kinematic
complexity. In last two decades it has been realized that collective
phenomena in shear flows are characterized by so-called non-modal
processes, which in turn are related with non-normality of the
involved mathematical operators \cite{tref}. In general, standard
stability theory, i.e.\ normal mode analysis, does not describe
completely the appearance of instabilities \cite{tref,crim,reddy}.
An alternative approach is based on the method developed by Lord
Kelvin \cite{kelvin}. In the framework of this method, the system of
partial differential equations governing the dynamical evolution of
modes of collective behavior is reduced to the inspection of a set
of ordinary differential equations (ODEs) in time, i.e., to the
solution of a relatively simple initial value problem.

This approach can be effectively used in (magneto-)hydrodynamics,
both for magnetized \cite{andro,chven} and unmagnetized plasma flows
\cite{rcb97}. In particular, in \cite{andro} the authors considered
the shear induced unstable modes applied to different, interesting
astrophysical prototype structures - jets. The problem was examined
for incompressible flows and it has been found that Alfv\'en waves
become subject to extremely strongly pronounced instabilities. It
was shown that the shear flow instabilities may lead to the
generation of large amplitude Alfv\'en waves. A similar problem, for
compressible magnetohydrodynamic flows, was studied by \cite{chven},
where the authors have argued that the flow inseparably blends the
slow and the fast magnetosonic and Alfv\'en modes, leading to an
efficient energy transfer from the background flow to the waves.

In \cite{rcb97} the authors examined the electrostatic perturbations
in an non-magnetized electron-ion plasma flow. {\it Unlike the
present paper}, the authors considered the problem for a simple,
one-dimensional, linearly sheared flow. It was shown that the
ion-sound waves turn into plasma oscillations caused by a very
efficient, shear-induced energy transfer mechanism between the mean
flow and the waves.

If the dissipation factors are efficient enough, the shear induced
instability might lead to a substantial heating of the plasma flows
\cite{rop10,orp12}. In particular, in \cite{rop10} the authors
considered acoustic waves and showed that the efficiency of the
so-called {\em self-heating} by acoustic wave perturbations might be
extremely efficient. The mentioned problem has been discussed in the
context of non-magnetic chromospheric heating in solar-type stars. A
very similar result has been obtained for magnetized flows
\cite{orp12}, where it was found that the rate of the self-heating
mechanism might be high enough to be related with realistic heating
scenarios in the solar atmosphere.

The scope of the present paper is to generalize the approach
developed in \cite{rcb97} and to study the shear flow dynamics for
ion electrostatic perturbations developing in more complex velocity
configurations. In particular, as we have already emphasized, in
previous studies the problem was considered for very simple,
one-dimensional velocity shears. On the other hand, it is well known
that in real astrophysical flows the kinematics  might be quite
complicated. Therefore, it is worthwhile to study more complex cases
and see how the generalization, i.e.\ the increased degree of the
flow complexity, alters the results qualitatively and/or
quantitatively. In the present paper, we examine the problem
physically, as a systematic plasma physics problem, without concrete
astrophysical applications. Further applications to different
astrophysical situations are in preparation and will be presented in
separate publications in due time.

The present paper is arranged in the following way. In the following
section, we develop the theory of shear-induced instabilities for
the ion perturbations. In the third section, we present and describe
our results, while in the final section we summarize them and
discuss them.

\section[]{Main Consideration}

In the present paper, we consider the non-magnetized, collisionless
plasma flow with an electron temperature much higher than that of
the ions, i.e.\ $T_e\gg T_i$. It is well-known that this kind of
plasma sustains weakly damped low-frequency longitudinal
electrostatic ion-sound waves with a constant ion acoustic speed
\begin{equation}
\label{Cs}  C_s = \sqrt{\frac{T_e}{m_i}},
\end{equation}
where $m_i$ denotes the ion mass.


We also assume that the quasi-neutrality condition holds, which
means that the equilibrium electric field equals zero. On the other
hand, the perturbations will inevitably lead to the generation of
the perturbed electrostatic field, ${\bf E}=-{\bf\nabla}\phi$, where
$\phi$ denotes the corresponding electric potential. Within the
limits of the low-frequency approximation, the electron number
density $N_e$ is governed by the Boltzmann distribution
\begin{equation}
\label{ne}  N_e = n_0exp\left(\frac{e\phi}{T_e}\right)\approx
n_0\left(1+\frac{e\phi}{T_e}\right).
\end{equation}
The other basic equations governing the evolution of the system
include the equation of mass conservation:
\begin{equation}
\label{con} D_t N_i + N_i \nabla \cdot {\bf  V}= 0,
\end{equation}
the momentum conservation equation:
\begin{equation}
\label{eul} D_t{\bf V} = -\frac{e}{m_i}{\bf \nabla}\phi,
\end{equation}
and the Poisson equation
\begin{equation}
\label{poi} \Delta\phi = 4\pi e\left(N_e-N_i\right),
\end{equation}
where $N_i$ denotes the ion number density, ${\bf  V}$ denotes the
flow velocity and $D_t \equiv
\partial_t + ({\bf V} \cdot \nabla)$
is the convective derivative.

In order to study the behavior of shear-induced instabilities, we
linearize the system of equations around the equilibrium state:
\begin{equation}
\label{n} N_{e,i} \equiv n_0 + n_{e,i},
\end{equation}
\begin{equation}
\label{v} {\bf V} \equiv {\bf U} + {\bf \upsilon},
\end{equation}
where by $n_{e,i}$ we denote the perturbed number densities of
electrons and ions, respectively, and ${\bf U}$ and ${\bf\upsilon}$
are the unperturbed and perturbed flow velocity components.
According to the linear approximation, it is assumed that all
perturbed quantities are much smaller than the corresponding
unperturbed quantities. After substituting Eqs.~(\ref{n},\ref{v})
into Eqs.~(\ref{con},\ref{eul}) they reduce to
\begin{equation}\label{con1}
\mathcal{D}_tn_i + N_i{\bf \nabla} \cdot {\bf\upsilon}= 0,
\end{equation}
and
\begin{equation}\label{eul1}
\mathcal{D}_t{\bf\upsilon} + ({\bf\upsilon} \cdot \nabla){\bf U} =
-\frac{e}{m_i}{\bf \nabla}\phi.
\end{equation}

In accordance with the method developed in \cite{mahand}, it is
assumed that the unperturbed flow velocity ${\bf U}$ is spatially
inhomogeneous. We then expand the velocity field in a Taylor series
around the point $A(x_0,y_0,z_0)$, preserving only the linear terms:
\begin{equation}\label{velexpand}
{\bf U}={\bf U}(A)+\sum_{j=1}^3\frac{\partial{\bf U}(A)}{\partial
x_j}(x_j-x_{j0}),\end{equation}
with $j=1,2,3$ and $x_j=(x,y,z)$.

One can straightforwardly show that the following ansatz
\begin{equation}\label{anzatz}
F(x,y,z,t)\equiv\hat{F}(t)e^{\psi_1-\psi_2},\end{equation} with
\begin{equation}\label{fi1}
\psi_1\equiv\sum_{j=1}^3{K_j}(t)x_j,\end{equation} and
\begin{equation}\label{fi2}
\psi_2\equiv\sum_{j=1}^3U_j(A)\int{K_j}(t)dt,\end{equation}
reduces the system of equations to a set of ordinary differential
equations \cite{mahand}. $F(x,y,z,t)$ denotes the physical
quantities $n,\upsilon_x,\upsilon_y,\upsilon_z$ and
$\hat{F}=\{\hat{n},\hat{\upsilon_x},\hat{\upsilon_y},\hat{\upsilon_z}\}$
are the corresponding terms depending on time. By $K_j(t)$ we denote
the wave vector components, which are obeying the following
differential equations \cite{mahand}:
\begin{equation}\label{dk}
{\bf \partial_{t}K}+ {\bf S^T} \cdot {\bf K}=0,\end{equation}
where ${\bf S^T}$ is a matrix transposed to the shear matrix ${\bf
S}$:
\begin{equation}\label{S}
 {\bf S} = \left(\begin{array}{ccc} U_{x,x} & U_{x,y} & U_{x,z}  \\
U_{y,x} & U_{y,y} & U_{y,z}  \\ U_{z,x} & U_{z,y} & U_{z,z} \\
\end{array} \right )\equiv \left(\begin{array}{ccc} A_{11} & A_{21} & A_{13}  \\
A_{21} & A_{22} & A_{23}  \\ A_{31} & A_{32} & A_{33} \\
\end{array} \right ).\end{equation}
where $U_{i,k}\equiv\partial U_{i}/\partial x_k$.


%

The plasma processes are described by
Eqs.~(\ref{poi},\ref{con1},\ref{eul1}). Taking into account
Eqs.~(\ref{anzatz},\ref{fi1},\ref{fi2}) and omitting the symbol
"$\wedge$", one can rewrite them in the following dimensionless
form:
\begin{equation}\label{cont}
d^{(1)}={\bf k} \cdot {\bf u},\end{equation}
\begin{equation}\label{ux}
u_x^{(1)}+ a_{11} u_x+a_{12}u_y+a_{13}u_z =
-k_x\varphi,\end{equation}
\begin{equation}\label{uy}
u_y^{(1)}+ a_{21}u_x +a_{22}u_y+a_{23}u_z =
-k_y\varphi,\end{equation}
\begin{equation}\label{uz}
u_z^{(1)}+ a_{31}u_x+a_{31}u_y+a_{31}u_z =
-k_z\varphi,\end{equation}
\begin{equation}\label{k}
\textbf{k}^{(1)} + {\bf s^{_T}} \cdot {\bf k} = 0, \end{equation}
\begin{equation}\label{S2}
 {\bf s} = \left(\begin{array}{ccc} a_{11} & a_{21} & a_{13}  \\
a_{21} & a_{22} & a_{23}  \\ a_{31} & a_{32} & a_{33} \\
\end{array} \right ),\end{equation}
\begin{equation}\label{pois1}
\varphi = \frac{d}{1+\frac{k^2}{\omega^2}},\end{equation}
where $d \equiv in_i/n_0$, ${\bf u} \equiv {\bf \upsilon}/C_s$,
$a_{ij}\equiv A_{ij}/K_z(0)C_s$ ($i,j = 1,2,3$),
$\varphi\equiv-iT\phi/e$, $\omega\equiv\omega_p/K_z(0)C_s$ and
$\omega_p=\sqrt{4\pi n_0e^2/m_i}$ corresponds to the ion-plasma
frequency. Moreover, $F^{(1)}$ denotes the derivative of the
function by a dimensionless time $\tau\equiv K_z(0)C_st$, and
$K_z(0)$ corresponds to the initial value of $K_z$.

We intend to study the flow dynamics from the point of view of shear
instabilities. For this purpose, in order to study the energy
transfer related with the perturbations, we introduce their total
energy:
\begin{equation}\label{etot}
E_{tot}\equiv \frac{{\bf u^2}}{2}+\frac{{d^2}}{2},\end{equation}
where the first and second terms are the kinetic energy  and the
compressional energy of the perturbations, respectively.

\section{Discussion}

In this section, we  study numerically different interesting regimes
related with the velocity-shear induced behavior of the
perturbations. One can examine the problem by considering two
different cases. Depending on the values of the shear matrix
parameters, the wave vectors either evolve exponentially, or exhibit
a stable character of evolution.

This can be easily seen for the particular case: $a_{13} = a_{23} =
a_{33} = 0$, $a_{22} = -a_{11}$, in which Eq.~(\ref{k}) simplifies
drastically so that $k_z$ becomes constant and the other two
components of the wave vector obey the following set of equations:
\begin{equation}\label{kx}
{k_x}^{(1)} -\Gamma^2k_x+A_x = 0, \end{equation} and
\begin{equation}\label{ky}
{k_y}^{(1)} -\Gamma^2k_y+A_y = 0, \end{equation}
where $\Gamma^2\equiv a_{11}^2+a_{12}a_{21}$, $A_x\equiv-a_{11}
a_{31}-a_{21}a_{32}$ and $A_y\equiv a_{11}a_{32}-a_{12}a_{31}$.

It is clear, that when $\Gamma^2>0$ the wave vector has an unstable
temporal evolution, varying exponentially in time. This in turn,
means that the physical system will inevitably undergo an unstable
behavior. On the other hand, if $\Gamma^2<0$ the wave vector will
depend on time periodically. But for specific values of the physical
parameters, even in this case, the instability may arise in spite of
the stable character of ${\bf k}(t)$.

\begin{figure}
  \resizebox{\hsize}{!}{\includegraphics[angle=0]{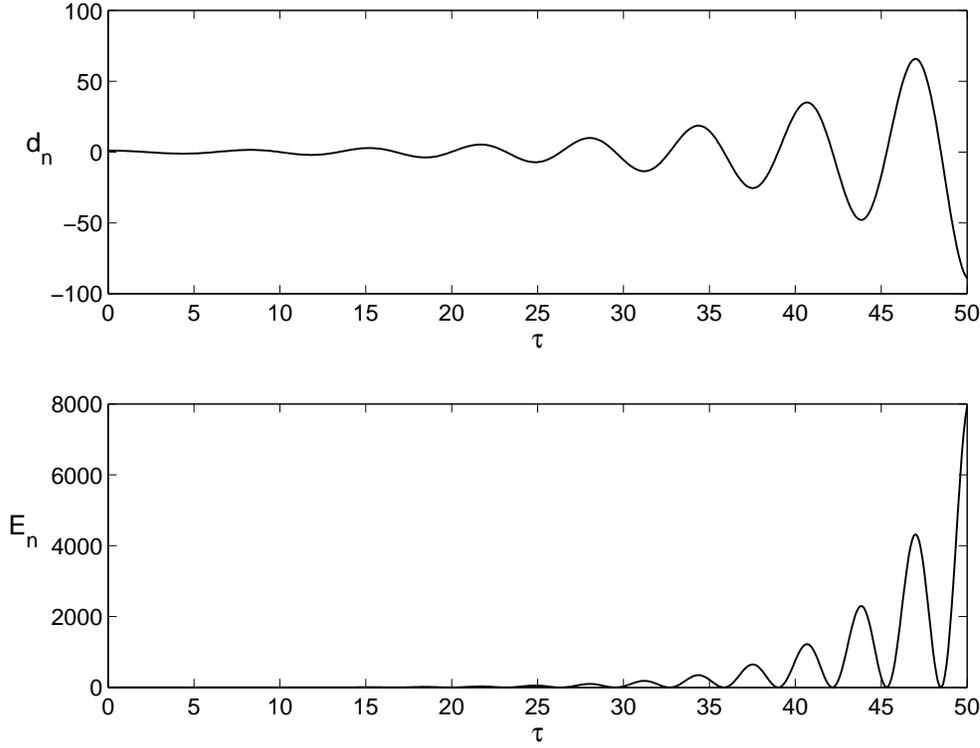}}
  \caption{The temporal behavior of the normalized density $d_n(\tau)\equiv d(\tau)/d(0)$ and energy $E_n(\tau)\equiv
  E_{tot}(\tau)/E_{tot}(0)$ perturbations. Here, the following set of parameters was chosen: $\omega = 1$, $a_{11} = -a_{22} = 0.1$,
$a_{12} = a_{13} =a_{21} =a_{23} =a_{31} =a_{32} =a_{33} = 0$,
$k_{x0}=k_{y0}=
  1$, $u_{x0} = u_{y0} = u_{z0} = 0$, $d_{0} = 0.1$.}\label{fig1}
\end{figure}

As a first example, we consider the situation when initially only
the density is perturbed. In Fig.~\ref{fig1} we show the temporal
evolution of the density and energy perturbations normalized by
their initial values. Here, the following set of parameters was
chosen: $\omega = 1$, $a_{11} = -a_{22} = 0.1$, $a_{12} = a_{13}
=a_{21} =a_{23} =a_{31} =a_{32} =a_{33} = 0$, $k_{x0}=k_{y0}= 1$,
$u_{x0} = u_{y0} = u_{z0} = 0$, $d_{0} = 0.1$, Therefore, $\Gamma^2
= 0.1$ implying that the wave vector varies in time exponentially.
In particular, as it is clear from the plots, for the considered
time interval, $\tau\in[0-50]$, the amplitude of the density
perturbations increases very rapidly from $1$ to $80$, which
consequently leads to an increase of the energy of perturbations up
to $\sim 8\times 10^3$. Generally speaking, this means that the
source of the amplification of the ion sound waves is the background
flow energy, which in the framework of the present approach, behaves
as an infinite energy reservoir.

\begin{figure}
  \resizebox{\hsize}{!}{\includegraphics[angle=0]{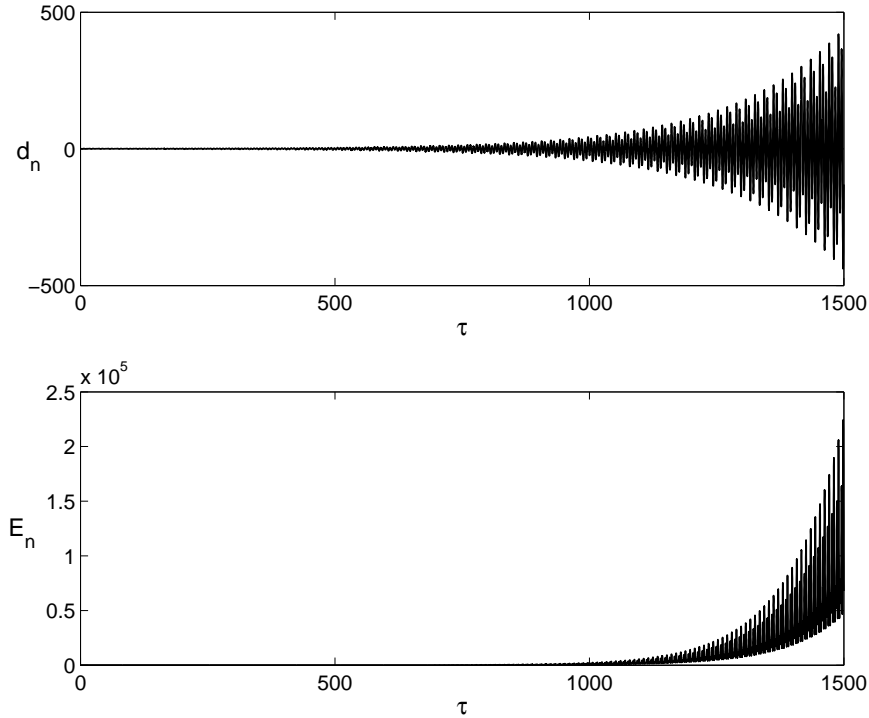}}
  \caption{The temporal behavior of normalized density $d_n(\tau)\equiv d(\tau)/d(0)$ and energy $E_n(\tau)\equiv
  E_{tot}(\tau)/E_{tot}(0)$ perturbations. Here, the following set of parameters was chosen: $\omega = 1$, $a_{12} = -0.594$, $a_{21} = 0.2$,
  $a_{11} = a_{13} =a_{22} =a_{23} =a_{31} =a_{32} =a_{33} = 0$, $k_{x0}=k_{y0}=
  1$, $u_{x0} = u_{y0} = u_{z0} = 0$, $d_{0} = 0.1$. Note that the range of $a_{12}$ where the evolution of the modes is parametrically
unstable, is very narrow, viz.\ [-0.607; -0.580].}\label{fig2}
\end{figure}
\begin{figure}
  \resizebox{\hsize}{!}{\includegraphics[angle=0]{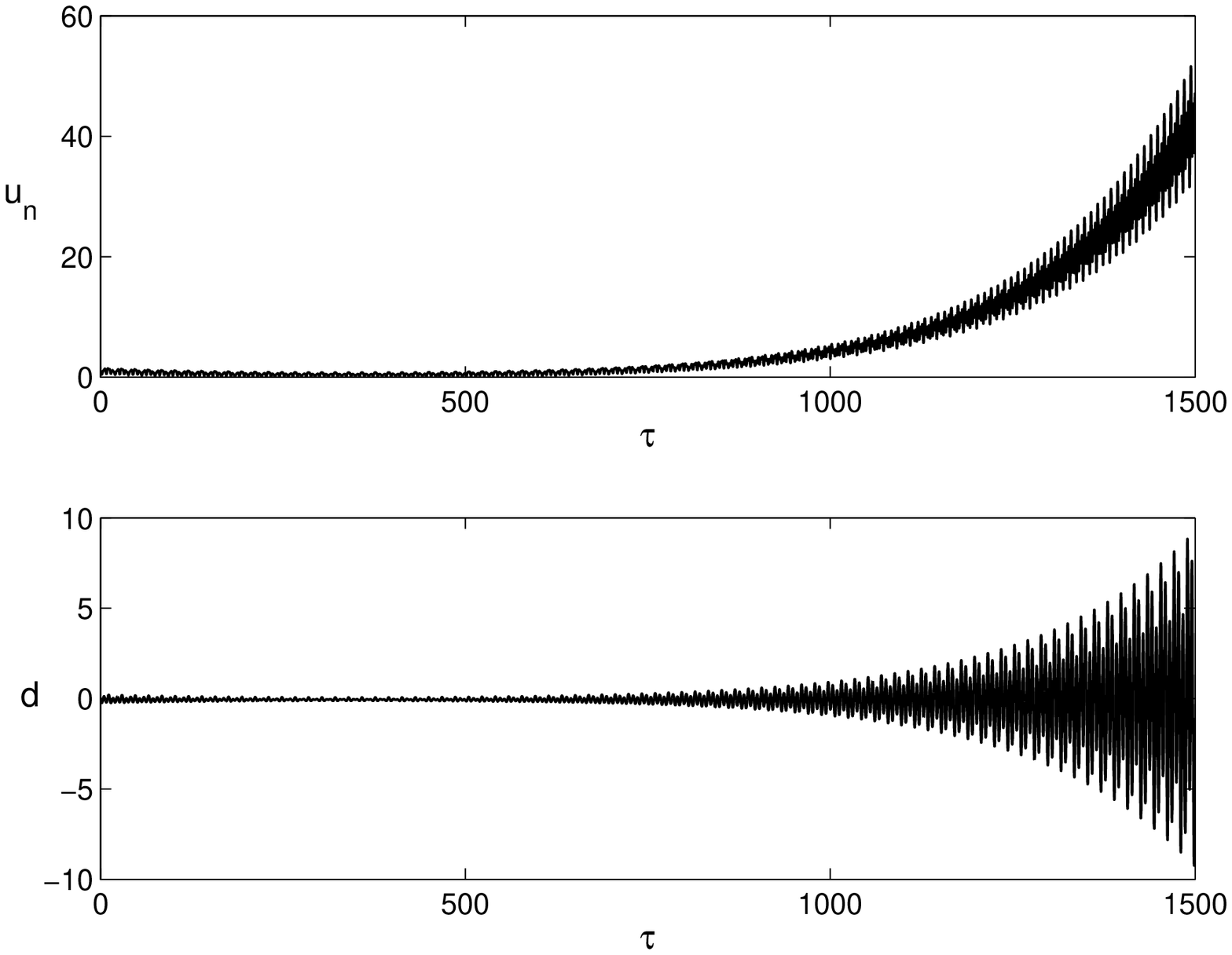}}
  \caption{The temporal behavior of the normalized total velocity $u_n(\tau)\equiv u(\tau)/u(0)$ and density $d(\tau)$ perturbations.
  Here, the following set of parameters was chosen: $\omega = 1$, $a_{11} = a_{13} = a_{22} = a_{23} = a_{31} = a_{32} = a_{33}$, $a_{12} = -0.594$,
  $a_{21} = 0.2$, $k_{x0}=k_{y0}=
  1$, $u_{x0} = 0.1$, $u_{y0} = u_{z0} = 0$, $d_{0} = 0$.}\label{fig3}
\end{figure}

However, as the investigation of the shear flow dynamics shows, the
system might undergo an extremely strong instability even for a
negative value of $\Gamma^2$. For studying this particular case, we
examine the following set of parameters: $a_{11} = -a_{22} = 0$,
$a_{12} = -0.594$, $a_{21} = 0.2$, while the rest of the parameter
values are the same as in the previous case. It is evident that now
$\Gamma^2<0$ and, hence, the wave vector is characterized by
periodic oscillations. In spite of this fact, the perturbations are
strongly unstable. In Fig.~\ref{fig2} we present again the behavior
of the density and energy perturbations. We see that the initially
perturbed sound waves amplify very rapidly. In particular, the
amplitudes of both the density and energy perturbations increase up
to $\sim 540$ and $\sim 2\times 10^5$, respectively. It is
worthwhile to note that this instability disappears if one slightly
changes the parameter values. As a matter of fact, for example, if
we change (increase or decrease) $a_{12}$ by no more than $\sim
2.4\%$, the system becomes stable. One can straightforwardly check
that the instability takes place only when $a_{12}\in[-0.607;
-0.580]$. One can thus conclude that this is, as we have
anticipated, an instability of parametric nature.

In the first two examples above, we examined the evolution of some
initially perturbed ion sound waves. These waves might be indirectly
excited by velocity perturbations. This particular example is
illustrated in Fig.~\ref{fig3}, where the time evolution of the
normalized total velocity, $u_n(\tau) =
\sqrt{u_x(\tau)^2+u_y(\tau)^2+u_z(\tau)^2}/u_0$ and density is
presented. Here, the following set of parameters was chosen: $\omega
= 1$, $a_{11} = a_{13} = a_{22} = a_{23} = a_{31} = a_{32} =
a_{33}$, $a_{12} = -0.594$, $a_{21} = 0.2$, $k_{x0}=k_{y0} = 1$,
$u_{x0} = 0.1$, $u_{y0} = u_{z0} = 0$, and $d_{0} = 0$. One can see
from the upper graph that the normalized total velocity is strongly
unstable, revealing the parametric instability. Even though
initially only the velocity is perturbed, we see that in due course
of time, density perturbations arise as well and, correspondingly,
parametrically unstable ion sound waves are generated.

\begin{figure}
  \resizebox{\hsize}{!}{\includegraphics[angle=0]{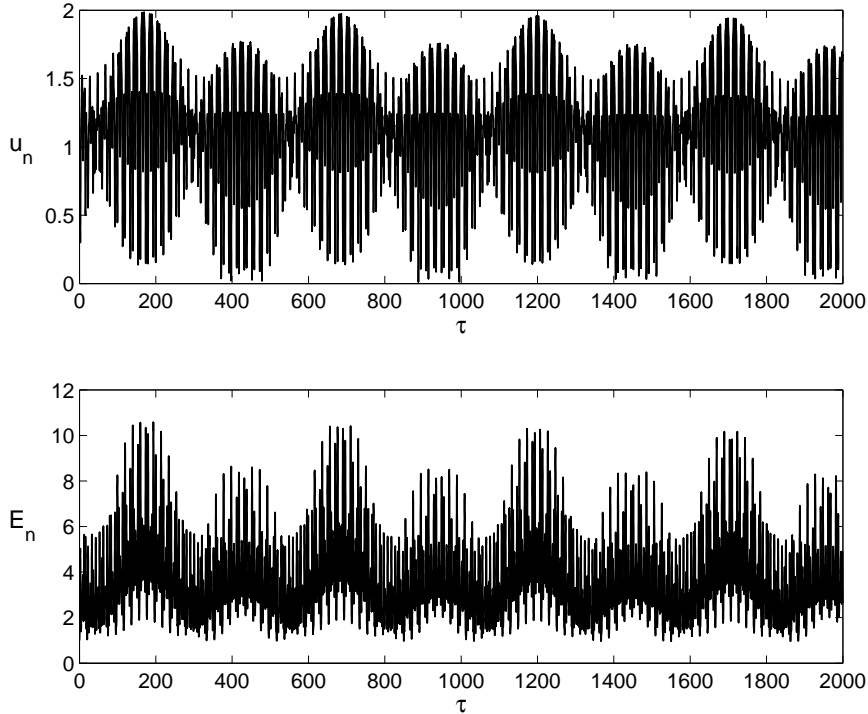}}
  \caption{The temporal behavior of normalized total velocity $u_n(\tau)\equiv u(\tau)/u(0)$ and energy $E_n(\tau)$ perturbations.
  Here, the following set of parameters was chosen: $\omega = 1$, $a_{11} = a_{13} = a_{22} = a_{23} = a_{32} = a_{33}$, $a_{12} =
-0.594$, $a_{21} = 0.2$, $a_{31} = -0.12$, $k_{x0}=k_{y0}=
  1$, $u_{x0} = 0.1$, $u_{y0} = u_{z0} = 0$, $d_{0} = 0$.}\label{fig4}
\end{figure}

It is well known that shear flows under favorable conditions exhibit
so-called beat modes \cite{rogm}. In Fig.~\ref{fig4}, we show the
behavior of both the normalized total velocity $u_n(\tau)\equiv
u(\tau)/u(0)$ and energy $E_n(\tau)$ perturbations. Here, the
following set of parameters was chosen: $\omega = 1$, $a_{11} =
a_{13} = a_{22} = a_{23} = a_{32} = a_{33}$, $a_{12} = -0.594$,
$a_{21} = 0.2$, $a_{31} = -0.12$, $k_{x0}=k_{y0}= 1$, $u_{x0} =
0.1$, $u_{y0} = u_{z0} = 0$, and $d_{0} = 0$. Figure~\ref{fig4}
displays an example of such a process. From this figure we see  that
the ion sound wave `pulsates', i.e.\ it is characterized by a
quasi-periodic temporal ``beat''. Such a structure is the result of
the superposition of two frequencies that differ only by a very
small amount. The fast oscillations are characterized by the ion
acoustic frequency, that is modulated by a lower frequency defined
by the shear parameters.

\section{Conclusions}

The goal of the present study was to consider ion-sound waves
influenced by {\it kinematic complexity} and to study the role of
the latter in the dynamical evolution of these waves. In particular,
in the framework of the method developed by \cite{kelvin} we have
considered the full set of equations, consisting of the momentum
conservation equation, the mass conservation equation and the
Poisson equation, and we linearized these around an equilibrium
state. We have shown that, depending on the choice of the set of
parameters, the system might undergo a very efficient instability.
In particular, it has been found that under favorable conditions,
due to the velocity shear, the wave vector becomes unstable, making
the ion-sound wave strongly unstable as well. On the other hand, we
have also shown an example of the excitation of acoustic modes by
means of only velocity perturbations. As yet another class of
instability, we have found that even when the wave vectors behave
periodically, for certain ranges of the parameter values, the system
becomes strongly unstable. Moreover, it has been shown that the
unstable character of the ion-sound waves dramatically changes to a
steady behavior  by  slightly changing the parameter values. Another
interesting feature of the examined system is that, under certain
conditions, it exhibits an "echo"-like behavior with quasi-periodic
pulsations of the ion-electrostatic modes.

In the near future, we plan to apply the developed method to
realistic astrophysical flow models. In particular, it would be
interesting to study the coupling of ion electrostatic waves with
the velocity shear in stellar atmospheres, including young stellar
objects. The idea that the ion-acoustic waves might influence the
properties of stellar atmospheres has a long-standing story. For
instance, in \cite{edwin} the authors have considered the ion-sound
waves in the solar atmosphere, studying the transition of flat
solitary waves into spherical modes. One of the intriguing problems
concerning solar physics is the so-called chromospheric heating,
that cannot be explained only by the convective component
\cite{carp}. In the framework of acoustic waves, there have been
proposed several mechanisms of chromospheric heating
\cite{lites,kalk}, but  observations with Transition Region And
Coronal Explorer (TRACE) NASA space telescope have shown that there
is about $90\%$ deficit in the energy flux required to heat the
chromosphere \cite{fossum_1,fossum_2,fossum_3}. Therefore, it is
interesting and quite reasonable to apply our model of kinematically
driven ion-sound waves to the solar chromosphere and study the
efficiency of heating.

In the present paper, we studied the fluid in a non magnetized
media, although in most of the astrophysical scenarios the magnetic
field plays a crucial role. Therefore, it is of fundamental
importance to generalize the present work by taking into account the
appearance of cyclotron modes and their possible coupling via the
agency of the velocity shear with ion-sound waves.

\section*{Acknowledgments}
The research of AR ad ZO was partially supported by the Shota
Rustaveli National Science Foundation grant (N31/49). ZO
acknowledges hospitality of Katholieke Universiteit Leuven during
his short term visit in 2012. AR also acknowledges partial financial
support by the BELSPO grant, making possible his visits to
CPA/K.U.Leuven in 2010-2012. SP acknowledges financial support of
the projects GOA/2009-009 (KU Leuven), G.0729.11 (FWO-Vlaanderen)
and C~90347 (ESA Prodex 9) in the framework of which the results
were obtained. The research leading to these results has also
received funding from the European Commission's Seventh Framework
Programme (FP7/2007-2013) under the grant agreements SOLSPANET
(project n° 269299, www.solspanet.eu) and eHeroes (project n°
284461, www.eheroes.eu).


\section*{References}
\begin{thebibliography}{100}
\bibitem{kerswell} Kerswell, R. R., 2005, Nonlinearity, 18, 17
\bibitem{schek} Schekochihin, A. A., Highcock, E. G. \& Cowley, S. C.,
2012, PPCF, 54, 055011

\bibitem{broder} Broderick, Avery E. \& Loeb, Abraham,
2009, ApJ, 703, 104L
\bibitem{kharb}
Kharb, P., Gabuzda, D. C., O'Dea, C. P., Shastri, P. \& Baum, S. A.,
2009, ApJ, 694, 1485
\bibitem{yso}
Chrysostomou, A., Bacciotti, F., Nisini, B., Ray, T. P., Eislöffel,
J., Davis, C. J. \& Takami, M., 2008, A\&A, 482, 575
\bibitem{pm98} Pike C.D., Mason
H.E., 1998, Sol. Phys., 182, 333
\bibitem{tref} Trefethen L.N., Trefethen A.E., Reddy S.C. Driscoll T.A., 1993,
Sience, 261, 578
\bibitem{crim} Criminale, W. O. \& Drazin, P. G., 1990, Stud. Appl. Maths., 83, 123
\bibitem{reddy} Reddy, S. C. \& Henningson, D. S., 1993, J. Fluid Mech., 252,
209
\bibitem{kelvin} Lord Kelvin (W. Thomson), 1887, Phil. Mag., 24,
Ser. 5, 188
\bibitem{mahand} Mahajan S.M., Rogava A.D., 1999, ApJ,
518, 814
\bibitem{rogm} Rogava A.D. \& Mahajan S.M., 1997, Phys. Rev. E., 55,
1185
\bibitem{andro} Rogava A.D., Mahajan S.M., Bodo G. \& Massaglia S., 2003, A\&A, 399,
421
\bibitem{chven} Rogava A.D., Bodo G., Massaglia S. \& Osmanov, Z., 2003, A\&A, 408,
401
\bibitem{rcb97} Rogava A. D., Chagelishvili, G. D. \& Berezhiani V.I.,
1997, Phys. Plasmas, 12, 4201
\bibitem{rop10} Rogava A.D., Osmanov Z. \& Poedts, S., 2010, MNRAS,
404, 224
\bibitem{orp12} Osmanov, Z., Rogava, A. D. \& Poedts, S.,
2012, Phys. Plasmas 19, 012901
\bibitem{edwin} Edwin P. M. \& Murawski K., 1995, Solar Physics, 158, 227
\bibitem{carp} Judge P. G. \& Carpenter K. G., 1998, ApJ, 494, 828
\bibitem{lites} Lites B. W., Rutten R. J. \& Kalkofen W., 1993, ApJ, 414, 345
\bibitem{kalk} Kalkofen W., 2007, ApJ, 671, 2154
\bibitem{fossum_1} Fossum A. \& Carlsson M., 2005, ApJ, 625, 556
\bibitem{fossum_2} Fossum A. \& Carlsson M., 2005, Nat, 435, 919
\bibitem{fossum_3} Fossum A. \& Carlsson M., 2006, ApJ, 646, 579




\end{thebibliography}
\end{document}